\documentstyle[preprint,aps]{revtex}
\begin{document}
\draft
\title{Optical-Model Description of Time-Reversal Violation \\
in Neutron-Nucleus Scattering}

\author{V. Hnizdo$^{1,*}$ and C. R. Gould$^{2}$}

\address{$^1$Department of Physics, Duke University, Durham, North Carolina
27708 \\
and Triangle Universities Nuclear Laboratory, Durham, North Carolina 27708 \\
$^2$Department of Physics, North Carolina State University, Raleigh,
 North Carolina 27695 \\
and Triangle Universities Nuclear Laboratory, Durham, North Carolina 27708}

\maketitle

\begin{abstract}
A time-reversal-violating spin-correlation coefficient in the total cross
section for polarized neutrons
incident on a tensor rank-2 polarized target is calculated by assuming
a time-reversal-noninvariant, parity-conserving ``five-fold" interaction in
the neutron-nucleus optical potential. Results are presented for the system
$n + {^{165}{\rm Ho}}$ for neutron incident energies covering the range
1--20 MeV. From existing experimental bounds, a strength of $2 \pm 10$ keV
is deduced for the real and
imaginary parts of the five-fold term, which implies an upper bound
of order $10^{-4}$ on the relative
$T$-odd  strength
when compared to the central real optical potential.
\end{abstract}

Measurements of the transmission of polarized neutrons through nuclear
targets provide sensitive tests of the fundamental symmetries in the
nuclear systems \cite{Bow90}. The violation of parity conservation in low
energy
neutron-nucleus scattering is now well established on the basis of such tests,
with measured longitudinal analyzing powers
of the order of 10\% \cite{Fra91}. Optical-model analyses
of the longitudinal analyzing power, utilizing a postulated
parity-nonconserving term in the neutron-nucleus interaction, have been
reported recently \cite{Koo92,Car93}. An optical-potential description
of nucleon-nucleus scattering observables supplies a useful analytical
tool as it relates the observables to the average properties
of compound
nuclear states, and provides a link to the underlying nucleon-nucleon
interaction \cite{Ger79}.

A neutron-transmission test of time-reversal invariance in neutron-nucleus
scattering
has been performed recently, employing polarized 2-MeV neutrons incident
on an aligned (tensor rank-2 polarized) $^{165}$Ho target \cite{Kos91,Kos92},
with a null result at a $10^{-4}$ level of a time-reversal-violating
(``$T$-odd") spin-correlation coefficient,  measured by
reversing the direction of neutron transverse
polarization. In the present work, we report
a coupled-channels calculation in the framework of the optical model of the
$T$-odd spin-correlation coefficient for the system $n+ {^{165}{\rm Ho}}$.

The calculation is based on the presence of
a time-reversal-noninvariant,
parity-conserving ``five-fold" term in the optical potential, expressed
in terms of the operator
${\bf s}\cdot({\bf I}\times{\bf\hat{r}})
({\bf I}\cdot{ \bf\hat{r}})$, where
${\bf s}$ and ${\bf I}$ are the projectile and target spins, respectively.
Unlike in the studies of parity nonconservation, where eV energies of $p$-wave
resonances are relevant and accordingly the compound-elastic cross section
dominates the shape-elastic (direct-elastic) cross section, MeV energies
with many overlapping, closely-spaced compound-nucleus resonances are under
consideration here. Following the general philosophy of the optical potential,
the five-fold term is an energy-averaged representation of
time-reversal-noninvariant  scattering processes, with an
imaginary part that accounts for time-reversal-noninvariant
contributions to the average compound-elastic and reaction cross sections. As
such it gives the most general description of time-reversal violation in the
scattering of polarized neutrons from aligned targets, and via folding-model
techniques \cite{Lov79}
it can in principle be related rigorously to a $T$-symmetry
violation in the effective nucleon-nucleon interaction.

The total cross section $\sigma_t$ for neutrons (spin $s=1/2$)
incident on a target nucleus with spin $I$, when the
projectile and target are in polarization states that
are described  by statistical tensors \cite{Sat83} that
are ``diagonal" in suitably chosen coordinate frames, i.e.
$\tilde{t}_{kq}(s)=\tilde{t}_{k0}(s)\delta_{q0}$ and
$\tilde{t}_{KQ}(I)=\tilde{t}_{K0}(I)\delta_{Q0}$, respectively,
can be written as
\begin{equation}
\sigma_t = \sum_{kK}\tilde{t}_{k0}(s)\tilde{t}_{K0}(I)\sigma_{kK},
\label{sigt}
\end{equation}
where
\begin{eqnarray}
\sigma_{kK}&=&
4\pi\lambdabar^2
\frac{\hat{k}\hat{K}}{\hat{s}\hat{I}}
{\rm Im}\sum_{\lambda}\hat{\lambda}\,
C_{kK\lambda}({\bf \hat {s}}\,{\bf\hat{I}}\,{\bf\hat{p}})
\sum_{Jljl'j'}(2J+1) \hat{l}\hat{\jmath}\hat{\jmath'} \nonumber \\
& & \times \langle l\lambda 00|l'0\rangle
W(JjIK;Ij')
{\mbox{$\left\{\!\!\begin{array}{ccc}
		l & s & j \\
		\lambda & k & K \\
		l' & s & j'
	    \end{array}\!\!\right\}$}}
T^J_{l'j',lj},
\label{sigkK}
\end{eqnarray}
with
\begin{equation}
C_{kK\lambda}({\bf\hat{s}}\,{\bf\hat{I}}\,{\bf\hat{p}})=
\frac{(4\pi)^{3/2}}{\hat{k}\hat{K}}
\left[[Y_k({\bf\hat{s}}),Y_K({\bf\hat{I}})]_
{\lambda},Y_{\lambda}({\bf\hat{p}})\right]_0.
\label{C}
\end{equation}
Here $\lambdabar$ is the reduced wavelength,
$\hat{k}=(2k+1)^{1/2}$ etc.,
$T^J_{l'j',lj} = (1/2i)(S^J_{l'j',lj} - \delta_{ll'}\delta_{jj'})$,
where
$S^J_{l'j',lj}$
are elements of the elastic-scattering $S$-matrix in the spin-orbit coupling
representation, $[\,,\,]_k$ denotes a spherical-tensor product of rank $k$,
and ${\bf\hat{s}}$, ${\bf\hat{I}}$, and ${\bf\hat{p}}$ are
unit vectors along the $z$-axes
of the frames in which the projectile and target statistical tensors are
diagonal, and along the beam direction, respectively;
the angular brackets, $W$ and  braces denote the
Clebsch-Gordan, Racah and 9-$j$ coefficients, respectively. The scalar
quantities
$C_{kK\lambda}({\bf\hat{s}}\,{\bf\hat{I}}\,{\bf\hat{p}})$
are the so-called correlation terms in the
forward elastic-scattering amplitude, which are
real (pure imaginary) for
$k+K+\lambda$ even (odd). Expressions of differing generality for the
total cross section with the projectile and target in polarization states
described by statistical tensors have been given also elsewhere
\cite{Alf73,Bar87,Hni87,Gou90}.

A correlation term $C_{kK\lambda}({\bf\hat{s}}\,{\bf\hat{I}}\,{\bf\hat{p}})$
indicates the presence of a term in the projectile-target interaction
that has the same  spherical-tensor structure.
For example, the $k=K=1$ cross section $\sigma_{11}$ has spin-spin correlation
terms with orbital angular momentum transfers $\lambda =0$ and 2:
\begin{equation}
 C_{11\lambda}({\bf\hat{s}}\,{\bf\hat{I}}\,{\bf\hat{p}}) =
 \left\{\begin{array}{ll}-\sqrt{\frac{1}{3}}\,{\bf\hat{s}}
\cdot {\bf\hat{I}}, & \lambda=0 \\
\sqrt{\frac{3}{2}}
\left[({\bf\hat{s}}\cdot {\bf\hat{p}})({\bf\hat{I}}\cdot{\bf\hat{p}})
- \frac{1}{3}\, {\bf\hat{s}}\cdot{\bf\hat{I}}\right],&\lambda=2
\end{array} \right.
\label{C11}
\end{equation}
and these reflect the presence of spherical and tensor spin-spin terms,
${\bf{s}} \cdot {\bf{I}}$ and $({\bf{s}}\cdot {\bf\hat{r}})
({\bf{I}}\cdot{\bf\hat{r}})
 -\frac{1}{3}\,
{\bf{s}}\cdot{\bf{I}}$,
respectively, in the projectile-target interaction (here ${\bf\hat{r}}$
is a unit
vector along the direction from the target to the projectile; an operator
quadratic in ${\bf\hat{r}}$ transfers two units of orbital angular
momentum, $\lambda=2$).
Another example is the case of $k=0$ and $K=2$, the deformation cross section
$\sigma_{02}$ for an unpolarized projectile incident on an aligned target.
The deformation cross section $\sigma_{02}$ has a  correlation
term with $\lambda=2$:
\begin{equation}
 C_{022}({\bf\hat{s}}\,{\bf\hat{I}}\,{\bf\hat{p}}) =
 \case{3}{2}\left[({\bf\hat{I}}\cdot{\bf\hat{p}})^2
-\case{1}{3}\right] = P_2({\bf\hat{I}}\cdot{\bf\hat{p}}),
\label{C02}
\end{equation}
which corresponds to the tensor potential $({\bf{I}}\cdot{\bf\hat{r}})^2-
\frac{1}{3}I(I+1)$ of a nucleus with spin $I > 1/2$, or the quadrupole
reorientation interaction of a rotational, statically deformed nucleus,
which has the same tensor form.

The $k=1$, $K=2$ cross section $\sigma_{12}$ is of our interest, as it
has a parity-even, time-reversal-odd
correlation term with $\lambda=2$, which has the following
``five-fold" form in terms of the Cartesian vectors ${\bf\hat{s}},\,
{\bf\hat{I}}$ and ${\bf\hat{p}}$:
\begin{equation}
 C_{122}({\bf\hat{s}}\,{\bf\hat{I}}\,{\bf\hat{p}}) =
i \sqrt{\case{3}{2}}\,{\bf\hat{s}}\cdot({\bf\hat{I}}\times{\bf\hat{p}})
({\bf\hat{I}}\cdot{ \bf\hat{p}}).
\label{C12}
\end{equation}
This correlation term is imaginary as here $k+K+\lambda$ is odd.
A projectile-target interaction that has
 the same spherical-tensor structure is
\begin{equation}
T_5=\case{1}{2}[{\bf{s}}\cdot({\bf{I}}\times {\bf\hat{r}})
({\bf{I}}\cdot{\bf\hat{r}})+
({\bf{I}}\cdot{\bf\hat{r}})({\bf{I}}\times{\bf\hat{r}})\cdot{\bf{s}}].
\label{T5}
\end{equation}
As ${\bf{I}}\times{\bf\hat{r}}$ and ${\bf{I}}\cdot{\bf\hat{r}}$ do not commute
(${\bf{s}}$ and ${\bf{I}}$ are {\it operators} of the projectile and target
spins, respectively, while the quantities
${\bf\hat{s}},\,
{\bf\hat{I}}$ and ${\bf\hat{p}}$ in the correlation terms are ``c-numbers"),
the five-fold operator $T_5$ is symmetrized as above. It is Hermitian and
conserves  parity, but it anticommutes with the operator of time
reversal. The operator $T_5$ generates an antisymmetric elastic-scattering
$S$-matrix, $S^J_{l'j',lj}=-S^J_{lj,l'j'}$, as it is odd on time reversal
and Hermitian; an operator $iT_5$, which is time-reversal-even and
anti-Hermitian, leads similarly to an antisymmetric $S$-matrix. This parallels
the behavior of the familiar central terms in the optical potential:
the central real part $V(r)$, which is
time-reversal-even and Hermitian, and the central imaginary part
$iW(r)$, which is time-reversal-odd and anti-Hermitian, both
generate a symmetric $S$-matrix \cite{Sat83}. Table 1 summarizes the
symmetry properties of these terms in the optical potential.
The presence of an interaction with the operator $T_5$ or $iT_5$ in the
neutron-nucleus optical potential leads to a nonzero cross section
$\sigma_{12}$, or a $T$-odd spin-correlation coefficient \cite{Con93} $A_5$,
defined as
\begin{equation}
A_5 =
\frac{\sigma^{\rm max}_{12}}{\sigma_{00}} =
\frac{1}{2\,{\bf\hat{s}}\cdot({\bf\hat{I}}\times{\bf\hat{p}})
({\bf\hat{I}}\cdot{\bf\hat{p}})}
\,\frac{\sigma_{12}}{\sigma_{00}},
\label{A5}
\end{equation}
where $\sigma^{\rm max}_{12}$ corresponds to the maximum value of the
five-fold correlation term
and $\sigma_{00}$ is the total cross section for an unpolarized beam and
target  (note that $A_5$ of this definition is by a factor of
$(15/32)^{1/2}$ smaller
 than the ``$T$-odd analyzing power" used in \cite{Kos91,Kos92}).
Experimentally, the $T$-odd spin-correlation coefficient
is determined from the  ratio
$(\sigma_\uparrow - \sigma_\downarrow)/2\sigma_{00}$, where $\sigma_\uparrow$
$(\sigma_\downarrow)$ is the total cross section for neutrons incident on
an aligned target  and polarized up (down) with respect to a direction
parallel to ${\bf\hat{I}}\times{\bf\hat{p}}$ \cite{Kos91,Kos92}.
It should be mentioned that an interaction of the same form
as in Eq.\ (\ref{T5}), but with the position operator ${\bf\hat{r}}$ replaced
by the momentum operator ${\bf{p}}$ has the same spherical-tensor structure
and symmetry properties as the operator $T_5$. However, such an interaction
has a second-order velocity dependence, as opposed to the static
character of $T_5$, and such interactions are generally considered to be of
much
less importance when a static interaction is available ({\it cf.} the case
of the tensor spin-spin interaction).

Using the techniques of spherical-tensor algebra \cite{Bri68}, the five-fold
operator $T_5$ can be expressed as a scalar product of rank-2 spherical
tensors
\begin{equation}
T_5=-i\sqrt{4\pi}\left[\, [Y_2({\bf\hat{r}}),{\bf{s}}]_2,
[{\bf{I}},{\bf{I}}]_2\,\right ]_0,
\label{TT5}
\end{equation}
and thus it is seen to be responsible for transfers $\lambda=2,
\,j_s=1$ and $j_I=2$ of the orbital, projectile-spin and target-spin
angular momenta, respectively, in the spin-orbit coupling representation
$\mbox{\boldmath $\lambda$}
+{\bf{j}}_s ={\bf{j}}_I$. The reduced matrix element
\cite{Bri68}
of $T_5$ is then proportional to
\begin{eqnarray} &&
-i\sqrt{4\pi} \langle s \| {\bf{s}}\| s\rangle \langle I \|
 [{\bf{I}},{\bf{I}}]_2 \| I\rangle \nonumber \\
 =&&-i\sqrt{\case{2\pi}{3}s(s+1)I(I+1)(2I+3)(2I-1)},
\label{RM}
\end{eqnarray}
and a calculation of elastic-scattering $S$-matrix elements with the
interaction $T_5$ included in the optical potential can be performed
using a standard coupled-channels code.
The  reduced matrix element of $T_5$
is imaginary, in accordance with the operator being odd on time reversal
and Hermitian.

Using the coupled-channels code CHUCK \cite{Kun}, calculations of the $T$-odd
spin-correlation coefficient
$A_5$ were performed for the system $n\, +\, ^{165}$Ho (spin $I=7/2$) for
neutron incident energies covering the range 1--20 MeV. A deformed
optical potential of the standard Woods-Saxon parametrization was employed,
with the real part of strength $49.8-16(N-Z)/A-0.325E$ MeV, surface imaginary
part of strength $5-8(N-Z)/A+0.51E\,(E\leq 6.5$ MeV) and
$8.3-8(N-Z)/A-0.09(E-6.5)\, (E > 6.5$ MeV), volume imaginary part of strength
$-1.8+0.2E\, (E > 6.5$ MeV), and a spin-orbit strength of 6 MeV; the reduced
radius and diffuseness parameters of all the terms of the potential were
1.26 and 0.63 fm, respectively, with the exception of 0.48 fm for the
diffuseness of the surface imaginary part; the central part of the potential
had a quadrupole deformation parameter $\beta_2 =0.29$ \cite{You}.
A five-fold interaction term
\begin{equation}
[V_5f_V(r)+ iW_5f_W(r)]T_5,
\label{VWT5}
\end{equation}
with volume Woods-Saxon form factors $f_V(r)$ and $f_W(r)$
of the same geometries as the corresponding
terms in the central potential was added to the optical potential.
Apart from the coupling due to the five-fold term, the calculations included
the reorientation coupling of the ground state of $^{165}$Ho, with angular
momentum transfers $\lambda =2,$ 4 and 6, assuming the rotational model
and a ground-state bandhead $K=I=7/2$.  Performing   coupled-channels
calculations, instead of a distorted-wave Born approximation treatment of
the small five-fold term, thus had the advantage of being able to account
easily for the large static deformation  of $^{165}$Ho.
The optical potential used is an adequate representation of the average
$n$-$^{165}$Ho interaction in the energy range considered;
this can be seen
in Figures 1 and 2, where the experimental total cross sections $\sigma_{00}$
and deformation cross sections $\sigma_{\rm def}= \sigma_{02}/P_2({\bf\hat{I}}
\cdot {\bf\hat{p}})$ are compared with the predictions calculated with the
potential. Figure 3 presents the results for the $T$-odd spin-correlation
coefficient $A_5$,
separately for a pure real five-fold term
of strength
$V_5=+0.1$ MeV and a pure imaginary five-fold term of strength $W_5=+0.1$ MeV.
As a function of the incident neutron energy,
the spin-correlation coefficients are seen
to oscillate in a typical Ramsauer fashion \cite{Gou86}, reflecting
the small changes
in the overall strength of the nucleon-nucleus interaction due
to the five-fold term.
The amplitude of the oscillation is
about $4 \times 10^{-3}$ at the low-energy end of the 1--20
MeV range. The calculated values of $A_5$ are proportional to the small
strengths $V_5$ and $W_5$.

Using these results and the experimental value of
$A_5 = (0.7 \pm 4.1)\times 10^{-4}$ for the $T$-odd spin-correlation
coefficient in the system $n + {^{165}{\rm Ho}}$ at 2 MeV \cite{Kos91},
both the real and imaginary strengths
of the $T$-odd five-fold term in the optical potential are estimated as
$2 \pm 10$ keV.
In order to be able to relate quantitatively this estimate to the strength
of the $T$-odd, parity-even part of an effective nucleon-nucleon interaction,
the five-fold term in the optical potential would have to be calculated
from the underlying nucleon-nucleon force, which is still an outstanding
task.
An order-of-magnitude estimate can be
made, however, of the bound on the ratio $\alpha_T$
of the strengths of the $T$-odd,
parity-even and $T$-even, parity-even parts of the
effective  nucleon-nucleon interaction
simply by taking
the ratio of the strengths of the five-fold and central real parts
in the optical potential: $\alpha_T < (10\,{\rm keV})/(50\,{\rm MeV})=
 2 \times 10^{-4}$. This is of the same order as the best sensitivity in
$\alpha_T$ obtained from analyses of detailed-balance experiments
\cite{Har90} and from a recent analysis of energy shifts in neutron
$p$-wave resonances \cite{Bar93}.

This work was supported in part by the US Department of Energy, Office of
High Energy and Nuclear Physics, under contracts DE-FG05-88-ER40441 and
DE-FG05-91-ER40619. V.~H. acknowledges support from the Foundation for
Research Development, South Africa.

\begin{figure}
\caption{The total cross section $\sigma_{00}$ for an unpolarized beam
and target for $n + {^{165}{\rm Ho}}$ as a function of the neutron incident
energy $E_n$. The experimental data are from \protect\cite{sig00}.}
\end{figure}
\begin{figure}
\caption{The deformation cross section $\sigma_{\rm def}$ for
$n + {^{165}{\rm Ho}}$ as a function of the neutron incident energy
$E_n$. The experimental data are from   \protect\cite{Fas73} (diamonds),
\protect\cite{Mar70} (squares) and \protect\cite{McC68} (circles).
Typical experimental errors are $\pm (50$--80) mb.}
\end{figure}
\begin{figure}
\caption{$T$-odd spin-correlation coefficient $A_5$
for $n+{^{165}}$Ho as a function of the
neutron incident energy $E_n$. The solid and dashed lines
are for a  real strength $V_5 =+0.1$ MeV and an imaginary strength
$W_5=+0.1$ MeV, respectively, of the five-fold interaction.}
\end{figure}
\begin{table}
\caption{The symmetry $S$ of the elastic-scattering $S$-matrix and
the properties under Hermitian conjugation $H$ and time reversal
$T$ for the central and five-fold interactions.
The positive (negative) sign denotes
that an interaction is even (odd) under a transformation and that the
corresponding $S$-matrix is symmetric (antisymmetric).}
\begin{tabular}{cccc}
Interaction  & $H$ & $T$ & $S$  \\
\tableline
$V(r)$       & $+$ & $+$ & $+$  \\
$iW(r)$      & $-$ & $-$ & $+$  \\
$T_5$        & $+$ & $-$ & $-$  \\
$iT_5$       & $-$ & $+$ & $-$
\end{tabular}
\end{table}
\end{document}